# Strain-induced polarization rotation in freestanding ferroelectric oxide membranes

*Alban Degezelle[1], Razvan Burcea[2], Pascale Gemeiner[2], Maxime Vallet[2,3], Brahim Dkhil[2], Stéphane Fusil[4], Vincent Garcia[4], Sylvia Matzen[1], Philippe Lecoeur[1], †Thomas Maroutian[1]

* alban.degezelle@universite-paris-saclay.fr

† thomas.maroutian@cnrs.fr

[1] Centre de Nanosciences et de Nanotechnologies, CNRS, Université Paris-Saclay, 91120 Palaiseau, France

[2] Université Paris-Saclay, CentraleSupélec, CNRS, Laboratoire SPMS, 91190 Gif-sur-Yvette, France

[3] Université Paris-Saclay, CentraleSupélec, ENS Paris-Saclay, CNRS, LMPS - Laboratoire de Mécanique Paris-Saclay, 91190 Gif-sur-Yvette, France

[4] Laboratoire Albert Fert, CNRS, Thales, Université Paris-Saclay, 91767 Palaiseau, France

**Abstract**
Freestanding ferroelectric membranes have emerged as a versatile tool for strain engineering, enabling the exploration of ferroelectric properties beyond traditional epitaxy. The resulting ferroelectric domain patterns stem from the balance at the local scale of several effects playing a key role, i.e. piezoelectricity linked to strain, and flexoelectricity arising from strain gradients. To weight their respective contributions for a given membrane geometry, the strain profile has to be mapped with respect to the ferroelectric polarization landscape, a necessary step to allow for a controlled tailoring of the latter. In this study, we examine the effect of bending strain on a Pb(Zr,Ti)O$_3$ membrane in a fold-like structure, observing a polarization rotation from out-of-plane to in-plane at the fold apex. Combining piezoresponse force microscopy, Raman spectroscopy, and scanning transmission electron microscopy, we map the ferroelectric polarization direction relative to the height profile of the membrane, and discuss the contributions of strain and strain gradients for this archetypal fold geometry. Our findings offer new insights into strain-engineered polarization configurations, and emphasize strain effects at the nanoscale to tune the functional properties in freestanding membranes.

# 1. Introduction

Ferroelectric freestanding oxide membranes have drawn a lot of attention for their wide range of potential applications in flexible electronics[1–3], ferroelectric topological structures generation[4,5], oxide-based devices on silicon[6,7], and many more. The fabrication processes are now maturing very fast, with major improvements achieved on membrane quality. Given the increasing number of available sacrificial layers[8–11], with different etchants and lattice parameters, the range of functional oxides that can be made into freestanding membranes is getting wider, making it a versatile technique for exploring physical properties in epitaxially grown thin films, albeit without substrate clamping.

One of the advantages offered by the membrane state of a material is that the morphology can be engineered to induce high strains and strain gradients, with orders of magnitude that can go far beyond the values achievable by epitaxy[12,13]. In the case of ferroelectric materials, the ferroelectric landscape is closely related to the strain landscape in a given layer. First, strain affects ferroelectric properties through the piezoelectric effect. In-plane tensile strain will typically favor the in-plane polarization direction, promoting polarization rotation from out-of-plane to in-plane upon biaxial or uniaxial strain application[14], as expected from piezoelectricity[15,16]. Second, the flexoelectricity[17] associated to strain gradients affects the spontaneous polarization by weakening or enhancing[13] it, depending on the gradient sign, magnitude and polarization orientation. Freestanding membranes give access to geometries that combine inhomogeneous strain and strain gradients, and the resulting interplay between strain / piezoelectricity and strain gradient / flexoelectricity opens up new ways to tune the ferroelectric landscape[4,18]. On one side, when the flexoelectric effect is large enough compared to the strain, it allows for a complete reversal of the polarization direction, as shown for instance with the strain gradient generated by the tip of an atomic force microscope[19,20]. The reported tip-induced ferroelectric polarization switches stem from a strong strain gradient ($\sim 10^6$ m$^{-1}$) obtained at a low strain level (<0.5%), suitable for a complete reversal of polarization. Bent membranes of the multiferroic $BiFeO_3$ also found their photoconductance affected by flexoelectric polarization[21] with gradients as low as 3 x 10$^4$ m$^{-1}$ and associated strain of +0.3%. On the opposite side, it is possible to harness the sole effect of strain and piezoelectricity to modify the ferroelectric state, such as achieved with homogeneous applied strain in stretching methods[22,23]. The behavior of ferroelectric materials subjected to both high inhomogeneous strain level and high strain gradient is however less documented, while it can lead to a variety of complex structures[5,12,13].

Controlling the ferroelectric polarization through on-demand strain and strain gradient patterns could trigger exotic polar textures and associated functionalities compared to the use of epitaxial misfit strain, for which the strain is limited by the available substrates and by epitaxy constraints. To pave the way for such flexible polarization engineering, one need to decipher the response of ferroelectric oxides in freestanding membranes with different types of strain landscapes.

In this study, we report on the experimental observation of polarization rotation in membranes of the archetypal ferroelectric $Pb(Zr,Ti)O_3$ having a fold-like morphology, with locally high strain levels (~ 1%) combined to strong strain gradients (~10$^5$ m$^{-1}$). Combining piezoresponse force microscopy (PFM), Raman spectroscopy, and scanning transmission electron microscopy (STEM), we map the effect of bending strain on the ferroelectric polarization in a membrane fold. We then discuss the origin of the induced polar landscapes in terms of the piezoelectric effect linked to the in-plane strain and of the flexoelectric effect linked to the strain gradient induced by the curved geometry. In particular, we evidence a polarization switch from out-of-plane to in-plane, beginning at low strain level (~0.2%) along the fold profile. This polarization is locked in-plane at the fold apex by the strong in-plane tensile strain, with no out-of-plane polarization reversal detected, despite a significant strain gradient that generates a flexoelectric polarization opposing the out-of-plane spontaneous polarization.

## 2. Results

### 2.1. Freestanding oxide membrane

A membrane comprised of a PbZr$_{0.2}$Ti$_{0.8}$O$_3$ (PZT) / SrRuO$_3$ (SRO) bilayer is fabricated from epitaxial films grown by pulsed laser deposition on a (001)-oriented SrTiO$_3$ (STO) substrate, using a 15 nm-thick La$_{2/3}$Sr$_{1/3}$MnO$_3$ sacrificial layer (**Figure 1a**). Details on thin film growth and membrane fabrication are given in the Methods section. X-ray diffraction confirms the epitaxial growth of the thin films on the STO substrate and the preserved crystalline structure of the bilayer membrane transferred on silicon (**Figure 1c**). The OOP lattice parameter of tetragonal (P4mm) PZT extracted from the (002) reflection is c$_{PZT}$ (as-grown) = 0.4110 nm for the as-grown epitaxial film, and c$_{PZT}$ (membrane) = 0.4114 nm for the membrane transferred on Si.

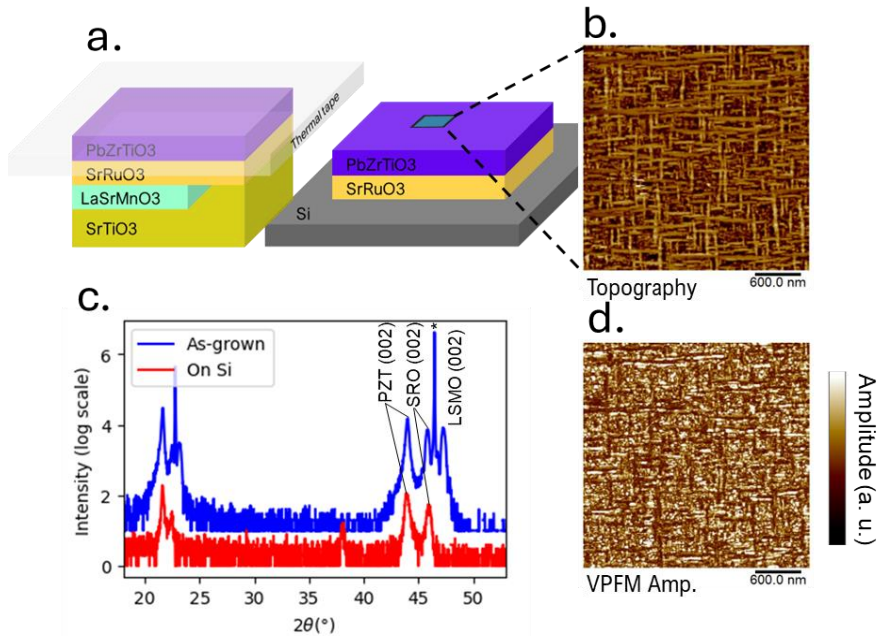

Figure 1: (a) Freestanding oxide membrane fabrication process: As-grown STO//LSMO/SRO/PZT stack with thermal tape on top, the LSMO sacrificial layer is dissolved to release the ferroelectric membrane. The released bilayer is then transferred to Si and the thermal tape removed with a heating process. (c) X-ray diffraction 2θ-θ scans of the as-grown films on SrTiO$_3$ substrate (blue), compared to the transferred membrane on Si (red), showing only PZT and SRO peaks on the latter. STO substrate (002) diffraction peak is marked by a star symbol. The weak peak around 38° is Ag (111) from silver paste contacting the membrane. (b) AFM image of the membrane surface, with 2 nm root-mean-squared roughness value, and (d) VPFM amplitude image acquired simultaneously, highlighting the *a/c* domain structure of the ferroelectric PZT membrane.

Atomic force microscopy (AFM) reveals a surface topography with a cross-hatched pattern (**Figure 1b**), comprised of alternating out-of-plane (OOP) *c*-axis oriented and in-plane (IP) *a*- axis oriented PZT domains, as confirmed by vertical PFM amplitude image (**Figure 1d**) showing an amplitude drop in the *a*-domains. This so-called *a/c* domain structure is typical for 130 nm-thick PZT film with this composition[16,24].

The Vertical PFM (VPFM) and Lateral PFM (LPFM) responses of the PZT for both the clamped and transferred flat states are displayed as Supporting Information (SI, **Figures S1 and S2**), as well as in **Figure 1** for the membrane state. While the same *a/c* domain structure is evidenced for both states, we note that the roughness is increased on the membrane after transfer, probably due to an orientation-dependent etching of the surface during the transfer process. Nevertheless, there is no noticeable difference in the PFM response and polar landscape of the PZT film before and after transfer.

**Figure 2a** shows the transferred PZT membrane on silicon observed by scanning electron microscopy (SEM). Various types of topographical structures such as folds, cracks, and rolls are evidenced alongside flat areas. Some of these topographical features can act as model structures to study the effects of strain on the physical properties of the membrane[25], such as rolls in PbTiO$_3$ / SrTiO$_3$ superlattices[26], or BaTiO$_3$ and BiFeO$_3$ folds[12,13]. They appear in an uncontrolled manner during the transfer process, due to the mechanical interplay between the thermal tape and the oxide membrane, together with strain relaxation upon releasing the latter from the substrate. We will focus in the following on the fold-type structure, to study the behavior of a 130 nm-thick PZT film under inhomogeneous bending strain.

## 2.2. Piezoresponse force microscopy

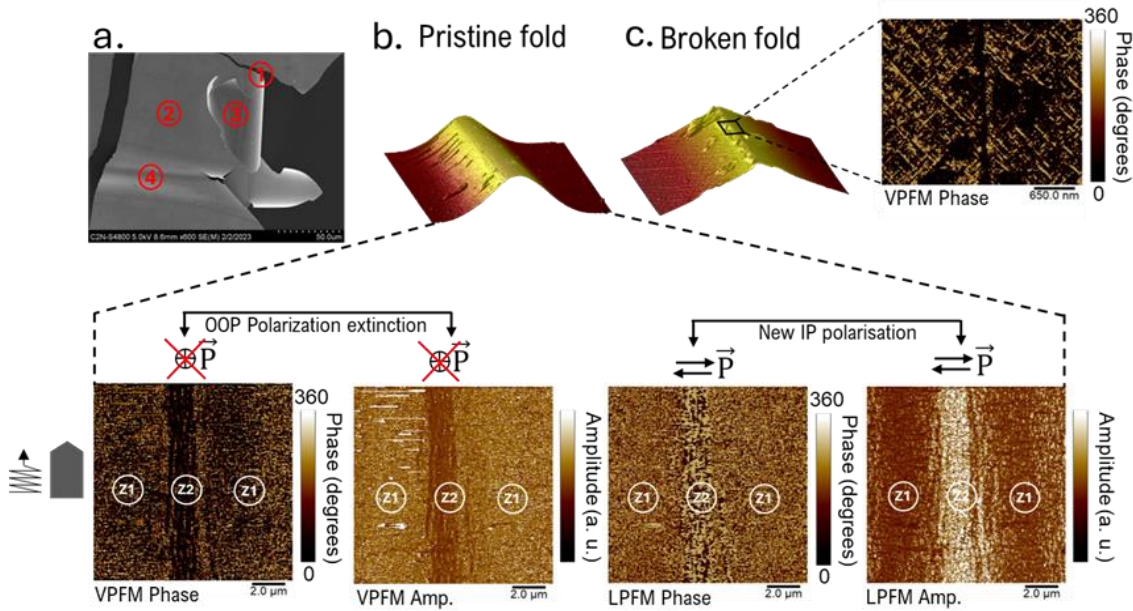

Figure 2: Ferroelectric domain structure on a typical fold of the PZT membrane. (a) SEM image of various topographical structures on the PZT membrane after transfer, such as (1) cracks, (2) flat areas, (3) rolls and (4) folds. (b) VPFM and LPFM phase and amplitude on the fold shown on top as 3D topography. Two different zones (Z1 and Z2) can be identified. There is an extinction of the as-grown out-of-plane polarization which has almost fully rotated to in-plane on top of the fold. Cantilever and scan orientations are indicated on the left side. (c) VPFM phase image on a broken fold, for which all strain is released in the film, as shown on the 3D topography. The pristine *a/c* domain structure of the flat areas is recovered everywhere on the structure. VPFM phase image corresponds to downward (dark) and upward (bright) contrasts respectively. In-plane phase image corresponds to polarization pointing to the right (dark) and left (bright) contrasts respectively.

Compared to the flat state described previously, PFM measurements on a PZT fold show a very different picture regarding the ferroelectric polarization landscape. Membrane folds are regions where the PZT thin film is fully freestanding above the Si substrate. In these structures, the PZT film is subjected to a bending strain depending on the surface curvature. To probe the effect of the local curvature on the polarization orientation, we carried out PFM scans on several PZT folds. **Figure 2** shows the VPFM and LPFM responses of a typical PZT fold. **Figure 2b** shows the 3D topography image of this fold. Few dots on the surface are polymer residues from the transfer process, giving spikes on the VPFM amplitude image. This fold reaches almost 850 nm in height, and its width (defined between the two points in contact with the Si) is about 12 μm.

The PFM images (**Figure 2b**) clearly show two different zones: The first zone (Z1) is where the response is the same as the flat state shown previously, with *a/c* domain structure. In the second zone (Z2), located on top of the fold, we observe an extinction of the OOP polarization, as emphasized by the drop of VPFM amplitude with respect to Z1. In the corresponding LPFM images, we observe in Z2 a strong LPFM amplitude signal coupled with a new IP domain structure evidenced in the LPFM phase signal. This Z2, with extinguished VPFM signal and enhanced LPFM, is about 2.5 μm wide. In Z1 on the LPFM amplitude, one can observe that the amplitude is lower on the ascending side of the fold just before entering the Z2. All these observations point to a 90° polarization rotation in the PZT membrane, with a sharp transition while going from the bottom to the top of the fold. The strong correspondence between the polarization arrangement and the fold profile hints that the observed rotation is tied to the bending strain in the film.

Strikingly, measurements on a broken fold with released strain in the PZT membrane displays the regular *a/c* pattern signature of the flat state, as shown through the VPFM phase on both sides of the broken fold (**Figure 2c**). This strongly supports that the fold-induced strain is at the origin of the observed polarization rotation in the pristine fold (**Figure 2b**).

To determine the orientation of the polarization in the IP domains at the top of the fold, the sample was rotated by 90° with respect to the cantilever to probe the orthogonal IP domain structure. The results of these scans are depicted in SI (**Figure S4**). The orthogonal scan ended up with no LPFM phase probed and zero LPFM amplitude detected, meaning that the IP ferroelectric polarization is indeed parallel to the profile direction, with no polarization component aligned to the fold axis. These results confirm the observation of the IP domain structure on top of the fold when the cantilever is parallel to the fold axis.

These PFM observations evidence a strain-induced polarization rotation in PZT membranes subjected to bending strain, from OOP to IP polarization. The ferroelectric domain structure is then no longer *a/c* on top of the fold, exhibiting a pure *a*-domain structure. The same observations were made on other folds, shown in SI (**Figure S8**). In order to confirm our PFM observations, and exclude any artificial effects coming from PFM tip mechanical force or driving voltage in such a peculiar folded configuration, we performed spatially-resolved Raman spectroscopy. This contactless optical method also discriminates *a*- from *c*-domains, limited to optical resolution but still suitable for the polar landscape under focus here.

## 2.3. Raman microscopy

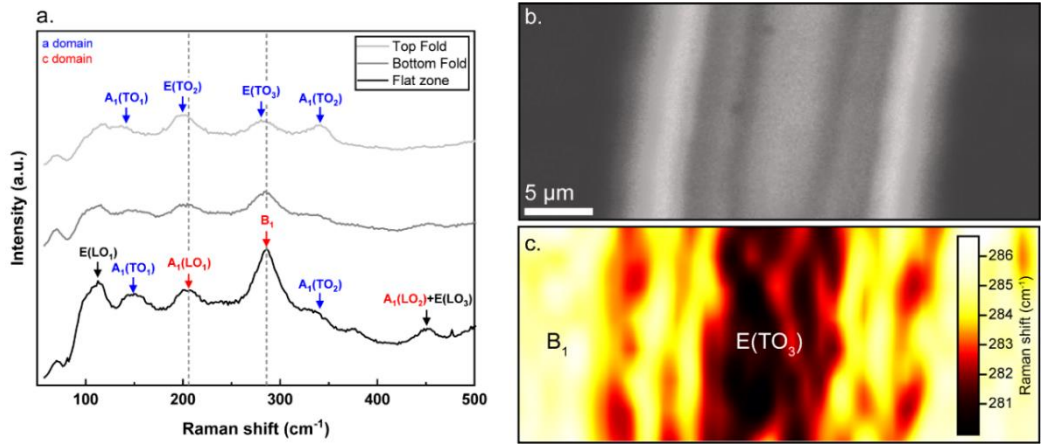

Figure 3: (a) Raman spectra of the PZT membrane acquired on a flat zone (black line), at the bottom (dark grey), and at the top (light grey) of the fold. The Raman modes associated with the *a*- and *c*- domains are labelled in blue and red, respectively. The dotted lines are aligned on *c*-domain modes and serve as a guide to the eyes. (b) Optical plane view of the fold recorded with the Raman spectrometer camera, and (c) corresponding Raman map showing the spatial distribution of the $B_1$ and $E(TO_3)$ modes, identified through their peak position indicated on the right.

Room temperature Raman spectra of the PZT membrane were acquired at different locations on the same fold as the one studied by PFM (**Figure 2**), and are displayed in **Figure 3** together with an optical plane view of the fold.

In the flat zone (black line) of a strain-free membrane, several modes are observed, including $A_1(LO)$, $B_1$, $A_1(TO)$, and $E(LO)$. The identification of phonon modes as a function of their symmetry is explained in SI (**Figure S5**). The $A_1(LO)$ and $B_1$ modes (red labels) are attributed to *c*-domains, while the $A_1(TO)$ modes (blue labels) correspond to *a*-domains, indicating the coexistence of these domains in this region. The presence of forbidden $E(LO)$ modes (black labels in **Figure 3a**) suggests that the domains are not perfectly aligned with the experimental setup. At the bottom of the fold (dark grey line), the coexistence of *c*- and *a*-domains persists, with a decreased intensity of some of the modes but no spectral shift. This is different from the spectrum obtained at the top of the fold (light grey line), for which a notable shift (highlighted by grey dashed lines) of the $A_1(LO_1)$ and $B_1$ modes is attributed to a transition to $E(TO_2)$ and $E(TO_3)$ modes, associated to *a*-domains. This shift implies that the domain population in this region predominantly consists of *a*-domains.

The Raman spectroscopy map in **Figure 3c** highlights the spatial distribution of the $B_1$ and $E(TO_3)$ modes, while going over the fold, as displayed in the corresponding optical plane view of **Figure 3b**. This map reveals a transition from $B_1$ to $E(TO_3)$ modes along the profile direction, while going toward the top of the fold. These results match our PFM data on the transition from *a/c*-domain to pure *a*-domain structure of the PZT along the fold profile.

## 2.4. Transverse electron microscopy

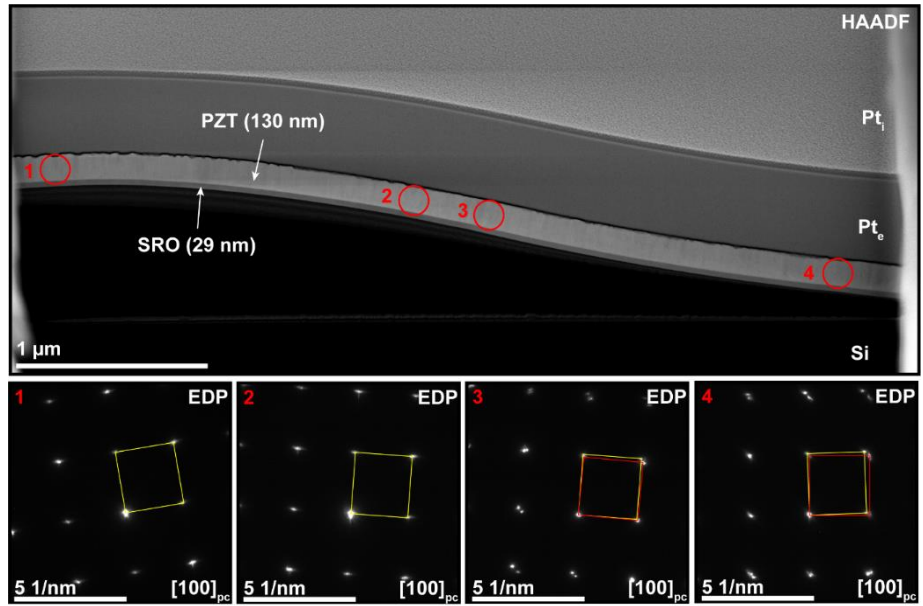

Figure 4: HAADF-STEM micrograph of the half-fold of the PZT. The red circles represent the SAED diaphragm used in four different areas of the fold: 1 at the top, 2 and 3 in the middle and 4 at the bottom. Below, the corresponding electron diffraction pattern (EDP) for each zone. Yellow and red rectangles represent the unit cells of $a$- and $c$-domain patterns, respectively.

**Figure** 4 shows a cross-sectional high-angle annular dark field (HAADF) STEM overview of the PZT half-fold structure, along the [100] pseudocubic (pc) zone axis for both oxide layers. The measured thickness of the PZT film is 130 nm, while the bottom electrode (SRO) exhibits a thickness of 29 nm. The fold height, defined as the distance between the silicon substrate surface and the bottom of the SRO layer, is approximately 700 nm, close to the one measured by AFM prior to the cross-section fabrication. The four marked circles indicate the specific regions where the diaphragm for selected-area electron diffraction (SAED) was employed to acquire the electron diffraction patterns (EDP) displayed below the STEM image.

The EDPs taken in the middle (3) and at the bottom (4) of the fold reveal a splitting of the diffraction spots, corresponding to two different unit cells of the reciprocal lattice, highlighted by yellow and red rectangles in **Figure 4**. The yellow rectangle having the higher length along [001] in reciprocal space is ascribed to $a$-domain PZT, while the red rectangle is assigned to $c$-domain PZT. The coexistence of $a$- and $c$- domains (4) persists up to the middle region (3) of the fold, while at area (2) the symmetry seems to abruptly transition to a pure $a$-domain configuration, which extends up to the top region (1). This transition occurs over a very narrow spatial range, less than 500 nm in width, consistent with the observations made by PFM (see **Figure 2**).

Bright field (BF) TEM micrographs of the same 1 (top) and 4 (bottom) fold areas are displayed in **Figure S6** (SI). Domain walls between $a$- and $c$- domains are clearly visible in (4), while no such features are visible in zone 1, thus confirming the observations made previously. All EDPs (**Figure 4** and **S6**) show some diffuse or streaky diffraction spots, stemming from local rotations and distortions of the PZT domains to accommodate the strain.

In order to estimate the strain level variation over the fold profile, it is possible to use the EDPs from **Figure 4**. First, the EDP scale was calibrated with respect to the Si substrate for all patterns. Second, as only *a*-domains are present over the whole PZT fold cross section, the strain was evaluated for these domains based on their associated unit cell (in yellow, **Figure 4**) dimensions in the different areas of the fold. PZT tetragonality is measured through the *c/a* ratio, which in the case of *a*-axis oriented PZT is given by $d_{010}/d_{001}$. Evaluating this ratio from the reciprocal unit cells of the EDPs gives a tetragonality of 1.06 and 1.05 at the top and at the bottom of the fold, respectively. With *a*-axis lying OOP, these values hint at an IP tensile strain of the order of 1% at the top of the fold with respect to the bottom. This result gives support to the mechanical modeling based on classical beam theory and the measured fold AFM height profile, as presented in the following section.

## 2.5. Mechanical model of bending strain in a membrane fold

In a fold-like structure, the freestanding membrane is subjected to inhomogeneous bending strain depending on the local curvature. Bending strain can be described by an IP strain, with the maximum tensile strain obtained at the surface of the top layer of the membrane, combined with a strain gradient along the film thickness. Based on classical beam theory, we calculated (see SI for details) both the IP strain along the profile of the fold and the different strain gradients in the film. We found that the most relevant one in this geometry is the IP strain gradient along the film thickness.

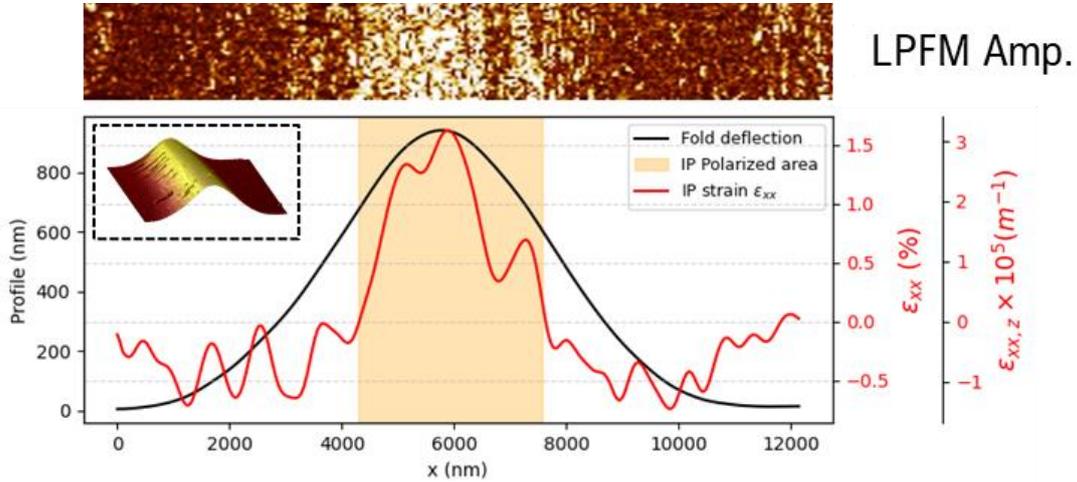

Figure 5: PFM height profile (black line) extracted from the fold shown in Figure 2b, together with the calculated in-plane strain (red line) from the bending strain model (see text). Note that the in-plane strain gradient has the exact same curve, albeit with a separate scale as indicated on the right. The shaded area marks the zone in which IP polarization is detected by PFM on the fold. The polarization orientation switch occurs at relatively low strain level on both sides of this zone. Top image is LPFM amplitude reproduced from **Figure 2**, aligned with respect to the *x*-axis of height and strain profiles. Inset shows the 3D topography of the fold. The height profile curve was slightly smoothed with a Gaussian filter ($\sigma = 4$).

**Figure 5** shows the height profile measured by PFM on the fold (from **Figure 2**), together with the corresponding calculated IP strain $\varepsilon_{xx}$ and associated IP strain gradient $\varepsilon_{xx,z}$ along the thickness direction (*z*), with *x* the fold profile direction. The calculated IP strain ranges from -0.5% compressive strain at the bottom of the fold up to +1.5% tensile strain at the top of the fold. **Figure 5** also shows the IPIP-polarized area detected by PFM on top of the fold, reproduced from **Figure 2**. We find that the polarization rotation between OOP and IP directions is happening at a tensile strain around +0.2% on both sides of the fold. The LPFM amplitude image arrayed on top of the graphs shows one large IP polarized stripe close to the center and a second, thinner stripe on the right. Strikingly, this pattern is recognizable on the strain curve in the form of one large bump peaking at +1.5% and a small bump peaking at +0.7%, which supports a direct link between IP strain and polarization switch. As a matter of fact, the slight change of slope is barely noticeable at the right of the fold height profile, while it gives a clear signature in the LPFM amplitude image, highlighting the second-derivative dependent local strain value. The same analysis done for another fold gives the same order of magnitude for the tensile train profile (SI, **Figure S8**), and evidences one more time a direct relationship between the IP strain level and the LPFM signal.

At this stage, considerations on the IP strain alone are able to match the observed polarization rotation from IP to OOP directions between bottom and top of a PZT fold. However, bending strain is characterized by IP strain combined to strain gradients along both the film thickness (*z*) and the profile direction (*x*), giving rise to flexoelectricity. We thus have to ascertain whether the latter can play a role in the observed polarization landscape of the PZT fold.

To estimate the relative orders of magnitude of the gradients in a membrane fold, we computed strain gradients for a Gaussian-shaped fold profile fitted to the experimental height profile (see SI for details). The calculated gradients are plotted in **Figure S7**, with $\varepsilon_{xx,z}$ the highest one (IP strain gradient along the film thickness) reaching about $3 \times 10^5$ m$^{-1}$ on top of the fold. A gradient of this magnitude has already been held responsible for modifying the ferroelectric state in a freestanding BiFeO$_3$ membrane on polymer[21], for a film thickness larger than 100 nm. In our case, since the corresponding flexoelectric coefficient of PZT is positive[27], this gradient induces a flexoelectric field pointing upward, in the same direction as the strain gradient. It is thus opposed to the as-grown polarization orientation in the *c*-domains of the film, where the polarization is pointing downward (as shown by PFM box-in-box writing, **Figure S9**). Still, no reversal of polarization is detected at the top of the fold, where the VPFM amplitude signal is minimal (**Figure 2**).

Gradients along the profile direction such as $\varepsilon_{zz,x}$ are maximum in the vicinity of the profile inflection points where the IP strain is vanishing (**Figure S7**). However, for the considered fold profiles, they are very low compared to those along the thickness direction such as $\varepsilon_{xx,z}$ and $\varepsilon_{zz,z}$ so their contribution to flexoelectricity will be very weak.

## 3. Discussion

It is well-known experimentally and theoretically[15,28] that lead-based oxide thin films exhibit different polarization domain structures depending on the misfit strain with various substrates. These domains are stable in a range of strain and temperature as described by Koukhar *et al.*[15], for homogenous misfit strain in an ideal epitaxial sample clamped on a substrate.

In the case of PbZr$_{0.2}$Ti$_{0.8}$O$_3$ epitaxial thin films, a bi-axial tensile misfit strain of ~0.35% is sufficient to stabilize a pure IP domain structure at room temperature[29]. On the other hand, experimental results with lead-based freestanding membranes have demonstrated that a large IP tensile strain of 4.6% via uniaxial stretching in a PbTiO$_3$ membrane stabilizes the IP polarization[14]. We estimated a noticeably smaller value, around 0.2%, for the switch of the OOP polarized *c*-domains to IP polarization in our PZT membrane folds. Note that our geometry does not impose bi-axial strain, but uniaxial strain along the profile direction, for a fully freestanding membrane above the Si substrate. Still, the IP strain can certainly be put forward as a main driving force behind the observed polarization rotation.

Now, we discuss why there is no OOP polarization detected on top of the fold, where a strong flexoelectric effect and associated OOP effective electric field are expected. Indeed, given the fold-like shape, at the top region of the fold both IP strain and OOP strain gradients are maximum. We estimate in our PZT membrane a strain gradient of about $3 \times 10^5$ m$^{-1}$ on top of the fold (**Figure 5**). In the literature, flexoelectric polarization switches from +*c* to -*c* domains via mechanical loading applied by an AFM tip were achieved via the flexoelectric effect[19,20,30]. Such a full reversal was observed with a strain gradient of $11 \times 10^5$ m$^{-1}$ in a 8 nm-thick PbZr$_{0.52}$Ti$_{0.48}$O$_3$ film[20], for a maximum IP strain of ~0.25%. On our PZT folds, the strain gradient is about 4 times smaller in an area where the IP strain is almost 6 times higher. Thus, the IP polarization is locked by the IP strain and no OOP polarization can be stabilized at the top of the fold. This is consistent with phase field simulations from Lu *et al.*[31], showing for PbTiO$_3$ films that a 1% biaxial tensile strain stabilizes the IP domain structure, and this even for large tip force (OOP strain gradient) applied over time. Finally, we note that this OOP flexoelectric field, opposite to the as-grown polarization, can promote its rotation toward the IP direction, thereby lowering the strain threshold to achieve a fully IP domain structure, in line with the low value of 0.2% estimated on our PZT folds.

## 4. Conclusion

In this work, we studied ferroelectric oxide membranes of PZT, focusing on their behavior in the vicinity of fold-like structures. Combining PFM measurements, with additional Raman spectroscopy and STEM imaging, we investigated the ferroelectric landscape of the PZT film subjected to the high bending strain level obtained in these folds. The induced IP strain can range continuously from -0.5% compressive to +1.5% tensile strain, which is beyond the typical misfit strain allowed by epitaxy while preserving high structural quality. We observe a polarization rotation in the membrane from OOP to IP, occurring at a low strain threshold (~0.2%). Considering the role of flexoelectricity for our PZT fold profiles, we suggest that the resulting polarization state is entirely driven by strain effects, despite a non-negligible strain gradient in the film. Our results give indications on how to pattern the polarization landscape in PZT or other ferroelectric materials by strain engineering in structures such as folds or periodic wrinkles.[4,32] By adjusting the membrane shape, deflection height, thickness and material composition, various polarization patterns could be generated, aiming at different polar textures. Advanced nanodevice such as nano-electro-mechanical system concepts could take advantage of finely balanced strain versus flexoelectricity, allowing to tune the coercive bending switching threshold, and the electrical detection of ferroelectric membrane curvature, deposited on any relevant flexible template to be monitored.

# 5. Methods

**Thin film growth:** The La$_{2/3}$Sr$_{1/3}$MnO$_3$ (LSMO, 15 nm thickness) / SrRuO$_3$ (SRO, 25 nm thickness) / Pb(Zr,Ti)O$_3$ (PZT, 130 nm thickness) multilayer structure was grown on SrTiO$_3$ (STO) substrate by pulsed laser deposition using a 248 nm KrF laser. The deposition was performed at a substrate temperature of 630°C, under a dynamic pressure of 0.15 mbar with N$_2$O gas for the PZT layer, and O$_2$ gas for the LSMO and SRO ones. The laser repetition rate was 5 Hz for SRO and PZT, 2 Hz for LSMO, at a common laser fluence of 2.5 J/cm$^2$ and a target-substrate distance of 5 cm. After growth, the sample was cooled down to room temperature under 400 mbar of O$_2$.

**Membrane fabrication:** In order to use the 15 nm-thick LSMO as sacrificial layer (SL), the sample is sonicated for several minutes in acetone and isopropyl alcohol, then a thermal release tape (TRT, Nitto ®) is applied to the sample surface to hold the freestanding membrane after SL dissolution. For the latter, the sample is immersed in a 200 mL etching solution of deionized water with 5 mL of HCl and 50 mg of KI. After two days, the film is released from the substrate and fully stuck on the TRT. Once the membrane is dry, it is stamped to a silicon substrate, and the sample is heated up first to 65°C for 5 min and then progressively (5°C / min) to 105°C, temperature at which the stickiness of the tape goes off.

**Piezoresponse force microscopy (PFM):** The measurements were conducted with a Bruker Dimension Icon atomic force microscope (AFM), using Pt-coated silicon tips (Multi75E, BudgetSensors). In order to probe the piezoresponse of the freestanding PZT membrane, AC voltage between 0.4 V and 1.2 V amplitude was applied to the SRO bottom electrode, with the tip grounded.

**X-ray diffraction (XRD):** XRD data were measured in parallel beam configuration with Cu K$\alpha_1$ radiation on a PANalytical X'Pert Pro diffractometer.

**Raman spectroscopy:** Measurements were carried out on a Labram Soleil Raman microscope from Horiba in backscattering mode with a x100 objective (spot size diameter of 0.5 µm), an excitation wavelength of 405 nm and a power of 0.13 mW. The grating was 2400 lines/mm, giving a spectral resolution of 1.5 cm$^{-1}$. Mapping of the region of interest (ROI) was acquired with a step size of 0.5 µm for X and 1 µm for Y over an area of 8x18 µm². As this device is not equipped with an analyzer, a second Raman spectrometer was used to discriminate the symmetry of the modes as a function of their position in cm$^{-1}$. This spectrometer is a T64000 from Horiba operating with a x100 objective and a wavelength of 488 nm. The instrument can be configured for (Vertical-Vertical) VV- and (Vertical-Horizontal) VH-polarization measurements. The identification of the modes is detailed in the Supporting Information.

**Scanning transmission electron microscopy (STEM):** The cross-section was prepared by focused ion beam on an FEI ThermoFisher Helios Nanolab 660, using a dedicated protocol to minimize the relaxation of strain/stress during the lamella thinning. This protocol was inspired from the work of Rivera *et al*.[33] STEM characterizations were performed on a FEI Titan3 G2 80-300 microscope, operated at 300 kV and equipped with a Cs probe corrector.

## Acknowledgements

We acknowledge support from the French Agence Nationale de la Recherche (ANR) through the FLEXO project (ANR-21-CE09-0046), and from the French national network RENATECH for nanofabrication. This work has received funding from the European Union's Horizon 2020 research and innovation programme under grant agreement No 964931 (TSAR project).

## Data availability statement

The data that support the findings of this study are available from the corresponding author upon reasonable request.

# Supporting Information

**Epitaxial PZT film before membrane fabrication**

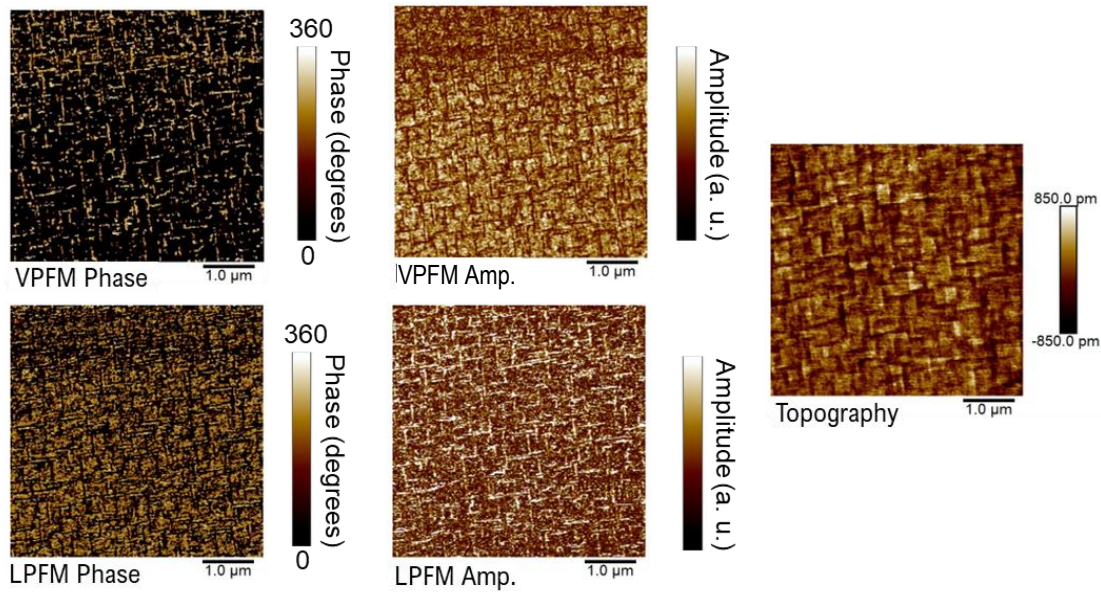

Figure S1: LPFM and VPFM scans of the as-grown STO/LSMO/SRO/PZT sample, and associated topography on right side with 0.2 nm root-mean-squared roughness value. Mixtures of $a$ and $c$ domains can be clearly observed from complementary out-of-plane and in-plane responses.

**Flat area on the PZT membrane**

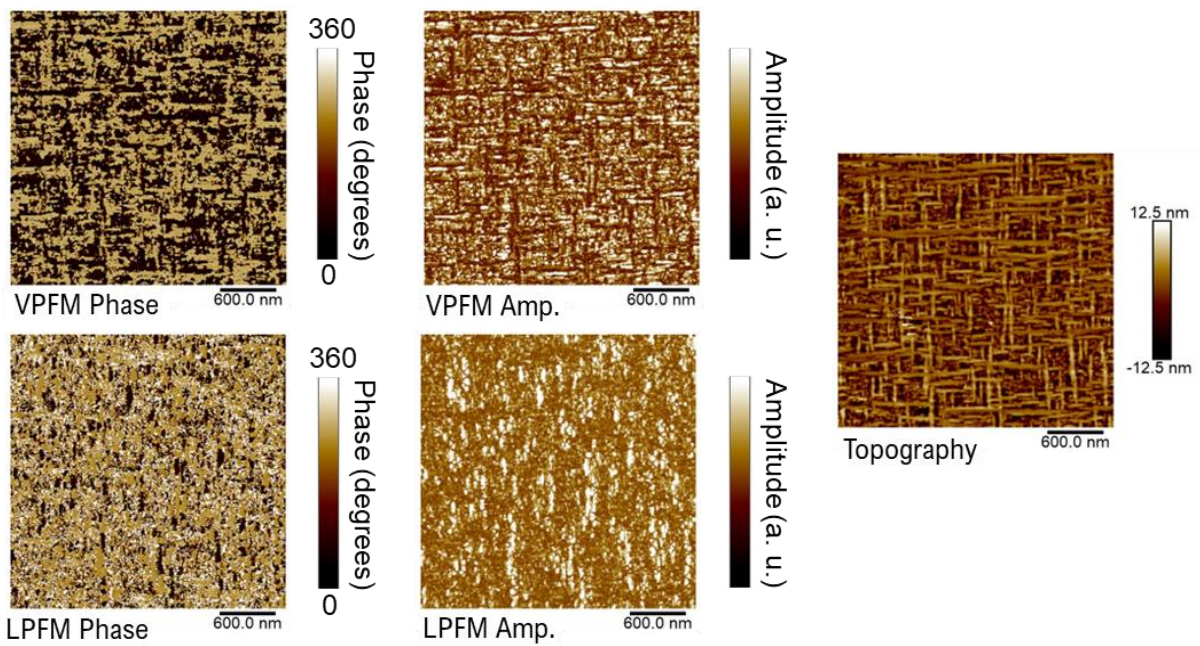

Figure S2: LPFM and VPFM scans of the transferred SRO/PZT membrane on silicon, and associated topography on right side with 2 nm root-mean-squared roughness value.

**Membrane transfer**

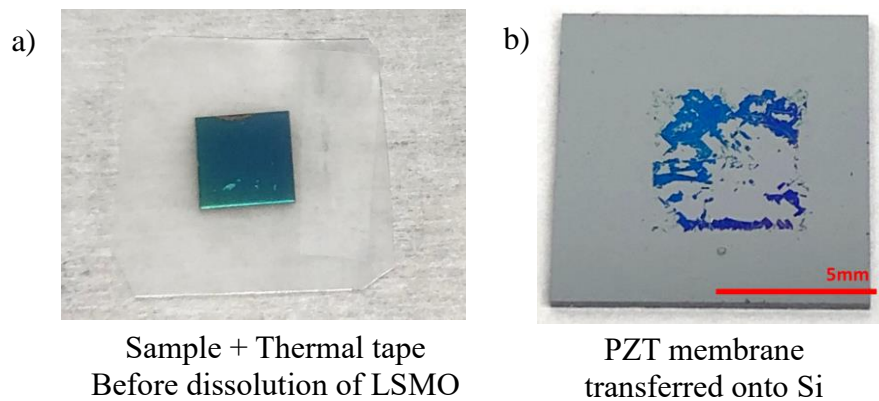

a) Sample + Thermal tape
Before dissolution of LSMO

b) PZT membrane
transferred onto Si

Figure S3: Optical images of the membrane (a) before and (b) after transfer on silicon.

**IP ferroelectric domain orientation on top of a PZT fold**

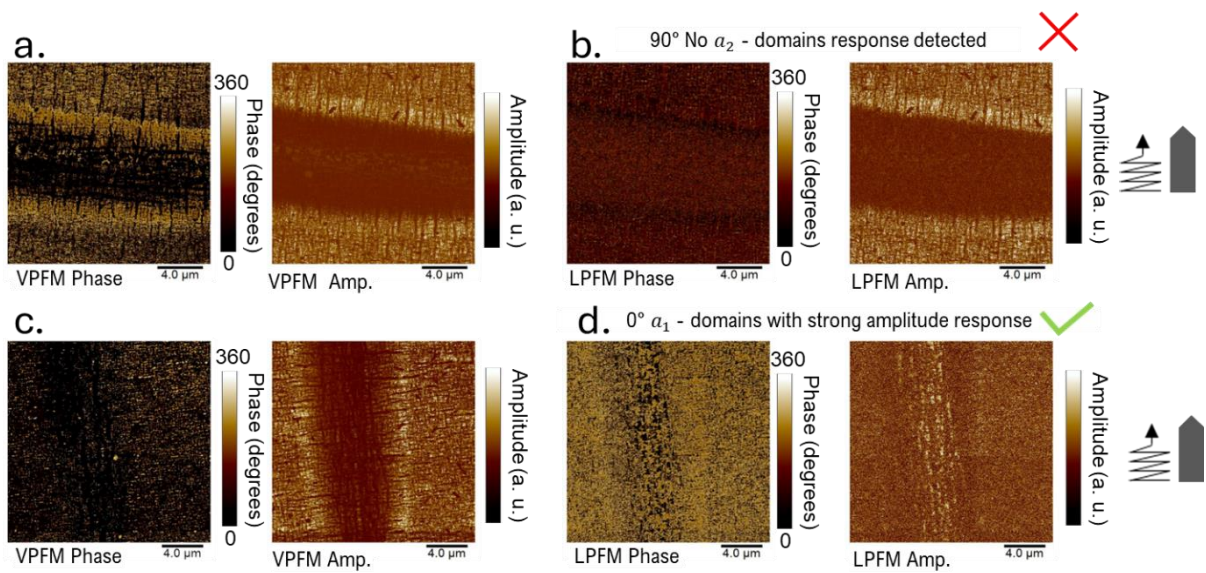

Figure S4: VPFM and LPFM scans at different cantilever orientations to probe orthogonal a-domains on a fold. VPFM at 90° (a) and LPFM at 90° (b). VPFM at 0° (c) and LPFM at 0° (d). No IP signal is detected on the fold while scanning at 90°, meaning the IP domains are polarized IP along the fold profile direction. Note that this is a different fold from the one shown in the main text.

# Raman spectroscopy

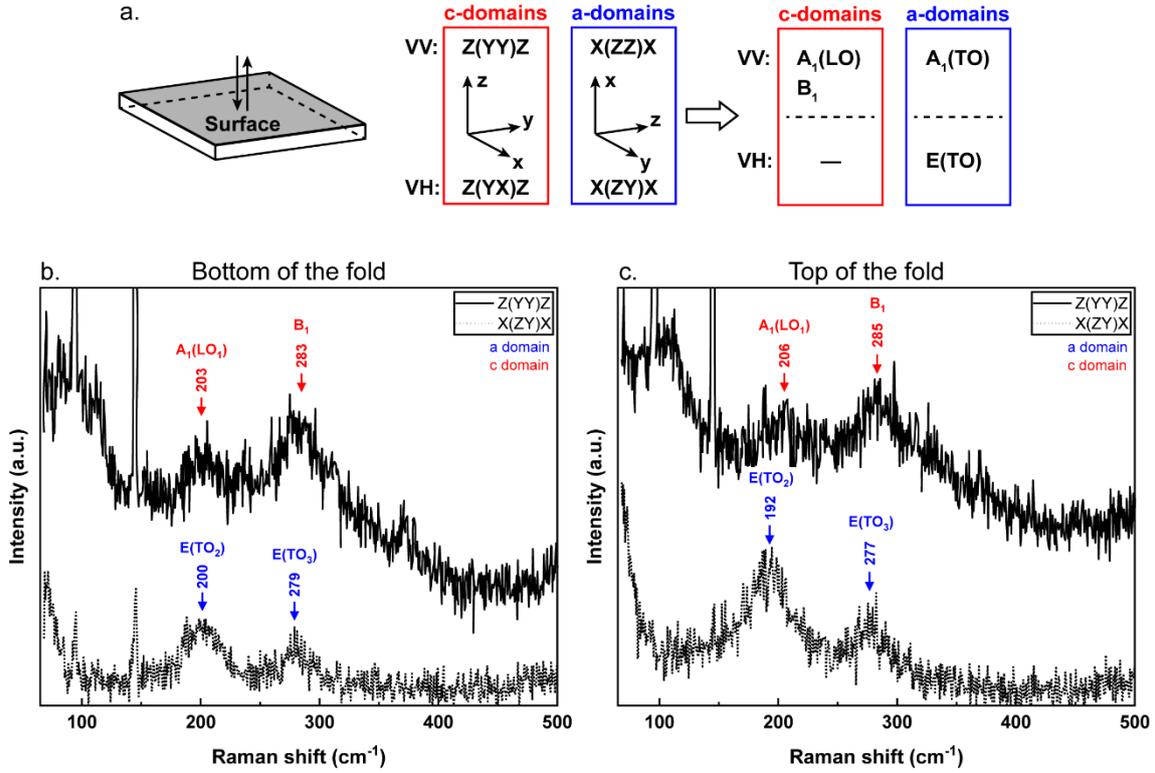

Figure S5: (a) Measurement configuration of the T64000 Raman spectrometer (Horiba) in VV- and VH-polarization conditions, with the corresponding selection rules and selected modes of *c*-axis and *a*-axis oriented PZT. (b) and (c) Out-of-plane and in-plane Raman spectra of the bottom and top of the fold respectively.

As described in the Methods section, the Raman spectrometer used to acquire the spectra of **Figure 3a** is not equipped with an analyzer, and therefore it is not possible to discriminate between the modes associated with *c*- or *a*- domains[34], i.e., OOP and IP modes, respectively. To address this limitation, a second Raman spectrometer was used, that allows measurements under VV- and VH-polarization conditions, in order to discriminate between IP and OOP domains.

**Figures S5b** and **S5c** show the OOP and IP Raman spectra at the bottom and top of the fold, respectively. According to Meng *et al*[35], PZT with a Zr/Ti ratio of 20/80 exhibits the same tetragonal symmetry as $PbTiO_3$ with coexistence of *a*- and *c*-domains. Based on this information and using the tabulated $PbTiO_3$ modes from Freire *et al*[36], the expected modes are depicted schematically in **Figure S5a** for each polarization configuration of the spectrometer.

The spectra obtained at the bottom and at the top of the fold are shown in **Figure S5b** and **S5c**, respectively. These measurements reveal that *a*- and *c*- domains are detected at both locations, albeit with some intensity fluctuations. The key point in these spectra is the frequency shift observed between the modes associated with the *c*- (higher frequency) and *a*- (lower frequency) domains. Knowing this, we can identify the peaks (modes) obtained with the spectrometer devoid of an analyzer (**Figure 3a**), showing a predominance of *a*-domains at the top of the fold and a coexistence of both domains at the bottom of the fold and in the so-called flat zone.

**Bright-field transverse electron microscopy**

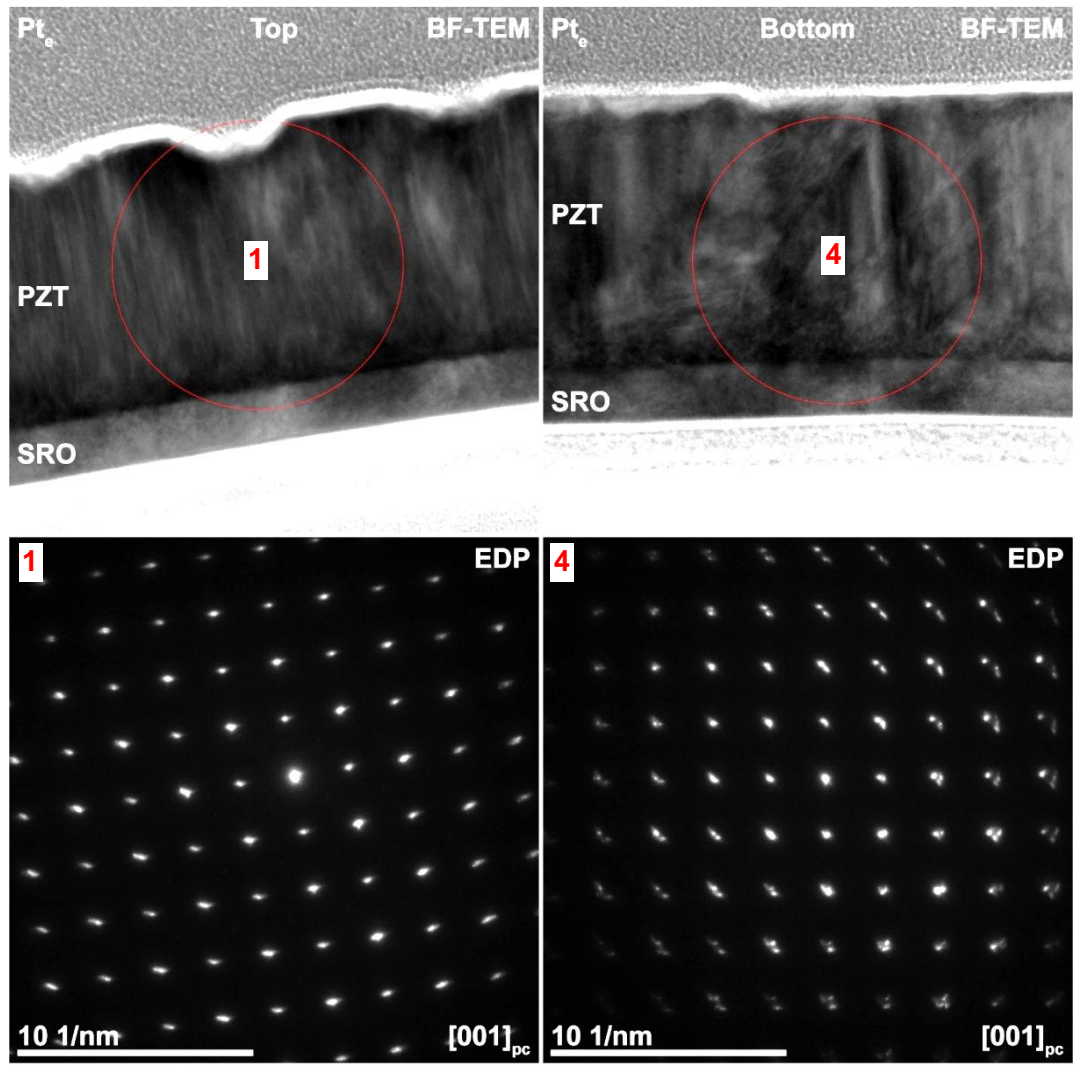

Figure S6: Bright Field (BF) TEM micrographs of the top and bottom of the PZT fold. The red circles represent the SAED diaphragms used to obtain the electron diffraction patterns displayed below.

**Figure S6** displays the Bright Field (BF) images of the top (1) and bottom (3) of the PZT fold. Below are shown the corresponding EDPs of zones 1 and 4. From a microstructural point of view, on the BF image of zone 4, it is possible to identify the domain walls between the *a*- and *c*-domains, seen as lines propagating along the thickness of the film. This feature is not visible on the BF image of zone 1, hence confirming that *a*- and *c*-domains coexist at the bottom of the fold, while *a*-domains prevail at the top. This observation is confirmed by the EDP obtained in each zone, here with a smaller camera length (thus a larger captured area of reciprocal space) than the one used for **Figure 4**, in order to see the higher order diffraction spots.

**Bending strain model**

To compute the IP strain $\varepsilon_{xx}(x,z)$ along the fold profile direction ($x$), we first calculate the neutral layer position relative to the SRO/PZT interface, defined at $z = 0$ ($z$ thickness direction). The PZT top surface is then at $z = +t_{PZT}$ and the SRO bottom surface at $z = -t_{SRO}$.

The position $\delta$ along the $z$-axis of the neutral layer is given by the following relation:

$$\delta = \frac{1}{2} \times \frac{(t_{PZT}^2 E_{PZT} - t_{SRO}^2 E_{SRO})}{(t_{PZT} E_{PZT} + t_{SRO} E_{SRO})} \quad (1)$$

Given the thicknesses $t_{PZT}$ = 130 nm and $t_{SRO}$ = 29 nm, and taking $E_{PZT}$ = 110 GPa[39] and $E_{SRO}$ = 120 GPa[37] as Young modulus for PZT and SRO layers, respectively, we get $\delta \sim +51$ nm. The neutral line is close to the middle of the PZT layer, due to a limited effect of the SRO one.

To get $\varepsilon_{xx}(x,\delta)$ at the PZT surface, we use the Euler-Bernoulli formula for a simple symmetrical bending of a beam with profile $w_g(x)$, the height profile of the fold:

$$\varepsilon_{xx}(x, \delta) = -(t_{PZT} - \delta)\frac{d^2 w_g(x)}{dx^2} \quad (2)$$

The OOP strain $\varepsilon_{zz}$ is obtained by means of the IP strain and the Poisson ratio of PZT ($\nu = 0.3$[38]):

$$\varepsilon_{zz} = -\nu \varepsilon_{xx}. \quad (3)$$

Expressions (1) and (2) were used to get the IP strain curves from the numerical derivative of the experimental fold height profiles of **Figure 5** and **Figure S8**. As the $\varepsilon_{xx,z}$ strain gradient is proportional to $\varepsilon_{xx}$ (see below), it was directly deduced from the latter.

However, in order to ascertain the relative weights of the flexoelectricity contributions, higher order derivatives of the height profile have to be considered. It is then advantageous to consider an analytical profile model, fitted to the experimental one, in order to calculate the relevant strain gradients for flexoelectricity.

The profile of the folds in our PZT membrane is best described by a Gaussian model:

$$w_g(x) = H e^{-\frac{(x-x_0)^2}{d_0^2}} \quad (4)$$

The model parameters can be readily obtained from fitting our AFM experimental data on a given fold. We note that cosine-based models are sometimes used to estimate the strain in bent membranes[39], but the Gaussian model gave the closest fit to the experiment in our case.

The expressions giving the relevant strain gradients are:

$$\varepsilon_{xx,z} = -\frac{d^2 w_g(x)}{dx^2} \quad (5)$$

$$\varepsilon_{xx,x} = -z\frac{d^3 w_g(x)}{dx^3} \quad (6)$$

$$\varepsilon_{zz,z} = \nu \frac{d^2 w_g(x)}{dx^2} \quad (7)$$

$$\varepsilon_{zz,x} = \nu z \frac{d^3 w_g(x)}{dx^3} \quad (8)$$

These strain gradients are represented in **Figure S7** for the Gaussian fold profile.

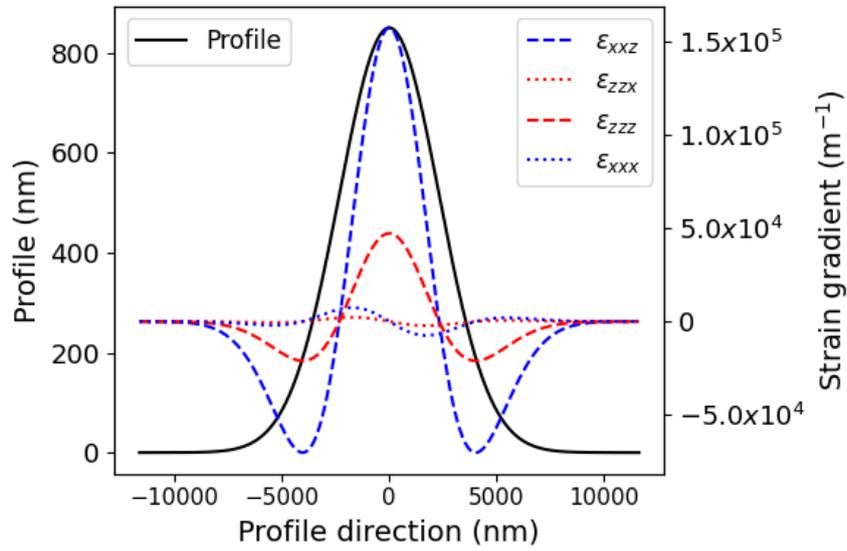

Figure S7: Strain gradients in a Gaussian-shape fold, as a function of the position on the profile of the fold. Dotted lines refer to gradients along the profile direction (x), and dashed lines to gradients along the film thickness (z). The flexoelectric contribution with effective electric field along the film thickness ($E_3$ along z), linked to $\varepsilon_{xx,z}$ and $\varepsilon_{zz,z}$ gradients, is the most important for this fold geometry.

We note that only 3 of these gradients are involved in flexoelectricity, through the flexoelectric fields $E_1$ and $E_3$ in $x$ and $z$ directions, respectively, defined by equations (9) and (10) with the corresponding flexoelectric coefficients $f_{xxzz}$, $f_{zzzz}$ and $f_{zzxx}$ :

| | | |
|---|---|---|
| $f_{xxzz}$ | 1.4 $\mu C/m$ | Ma and Cross[27] |
| $f_{zzzz}$ | 1.4 $\mu C/m$ | No reference, we used the same as $f_{zzxx}$ by default |
| $f_{zzxx}$ | 1.4 $\mu C/m$ | Ma and Cross[27] |

$$E_1 = \frac{f_{xxzz}}{\varepsilon_{xx}} \varepsilon_{zz,x} \quad (9)$$

$$E_3 = \frac{f_{zzzz}}{\varepsilon_{zz}} \varepsilon_{zz,z} + \frac{f_{zzxx}}{\varepsilon_{zz}} \varepsilon_{xx,z} \quad (10)$$

**Another fold on the PZT membrane**

In **Figure S8**, the LPFM amplitude signal overlaid on the height profile and IP strain curve for a different fold than the one displayed in **Figure 2c**. As the tensile strain in the membrane increases, the IP signal amplitude also increases, evidencing a direct relationship between the two. We can also note that in compressively strained regions, the LPFM amplitude is lowered.

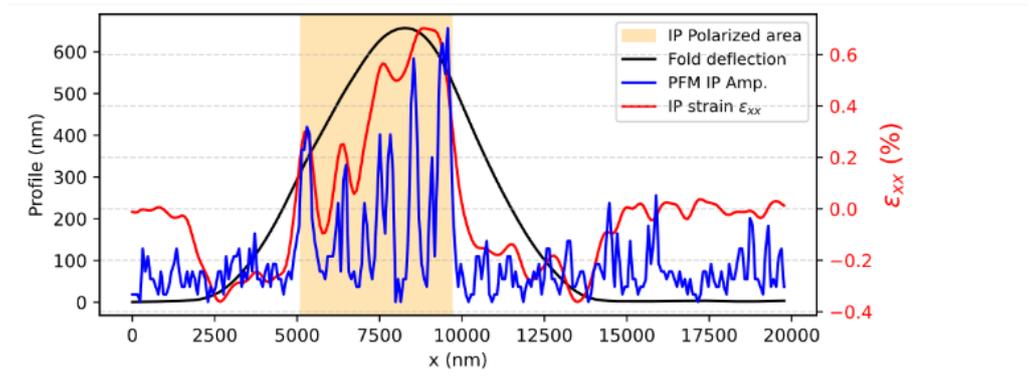

Figure S8: PFM height profile (black line, Gaussian filter σ = 1.5) extracted from a different fold than the one shown in Figure 2b, together with the calculated IP strain (red line) from the bending strain model, and the LPFM amplitude (blue curve) acquired together with the height profile. The shaded area marks the zone in which IP polarization is detected by PFM on the fold. Note that the LPFM amplitude signal closely follows that of the IP strain.

**Box-in-box PFM DC writing test on a flat zone of the membrane**

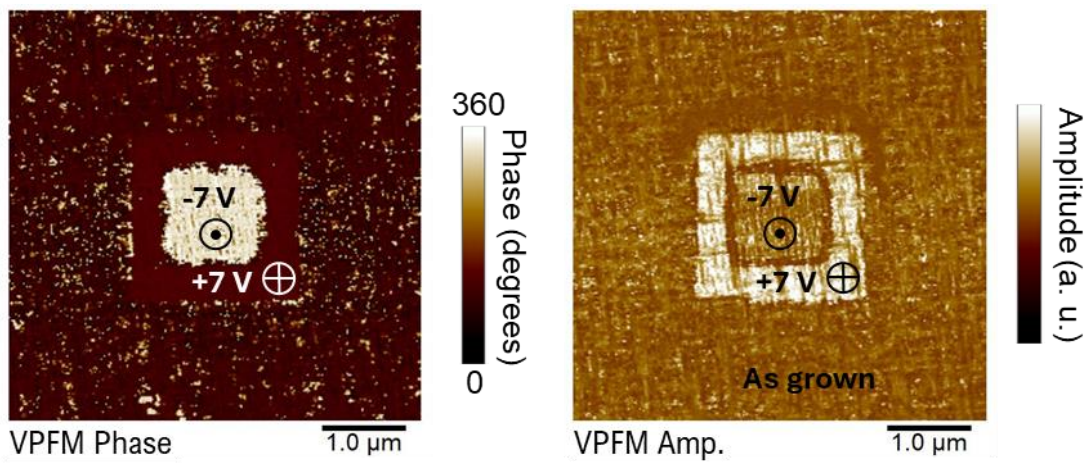

Figure S9: Box-in-box PFM DC writing via PFM on the flat transferred membrane, with DC voltage applied to the tip and sample grounded.